\documentclass[prb,prd,preprintnumbers,amsmath,amssymb,nofootinbib]{revtex4}

\usepackage{graphicx}
\usepackage{dcolumn}% Align table columns on decimal point
\usepackage{bm}% bold math
\usepackage{color}
\newcommand{\be}{\begin{equation}}
\newcommand{\ee}{\end{equation}}
\newcommand{\bea}{\begin{eqnarray}}
\newcommand{\eea}{\end{eqnarray}}

\newcommand{\bi}{\begin{itemize}}
\newcommand{\ei}{\end{itemize}}
\newcommand{\benu}{\begin{enumerate}}
\newcommand{\eenu}{\end{enumerate}}
\newcommand{\nn}{\nonumber}

\def \psl {p \kern-.45em{/}}
\def \qsl {q \kern-.45em{/}}
\def \ksl {k \kern-.55em{/}}
\def \lsl {l \kern-.40em{/}}
\def \ssl {s \kern-.45em{/}}
\def \asl {a \kern-.45em{/}}
\def \bsl {b \kern-.45em{/}}
\begin{document}

%\preprint{YNU-HEPTh-20-0220} 

\title{Electroweak fermion triangle loop contributions\\ to the muon anomalous magnetic moment revisited
 \vspace{0.5cm}}% Force line breaks with \\

\vspace{4cm}
\author{Ken SASAKI}
\email{sasaki@ynu.ac.jp}
\affiliation{Dept. of Physics, Faculty of Engineering,  Yokohama National University,  
 Yokohama 240-8501, JAPAN 
\vspace{0.7cm}}%

\begin{abstract}
The contribution to the muon anomalous magnetic moment from the fermion 
triangle loop diagrams connected to the muon line by a photon and a $Z$ boson is 
reanalyzed in the unitary gauge. With use of the anomalous axial-vector Ward identity, 
it is shown that the calculation in the unitary gauge exactly coincides with the one in  
the 't Hooft-Feynman gauge. The part which arises from the ordinary axial-vector Ward identity 
corresponds to the contribution of the neutral Goldstone boson. For the top-quark 
contribution, the one-parameter integral form is obtained up to  the order of $m_\mu^2/m_Z^2$. The results are compared with those obtained by the asymptotic expansion method.

\end{abstract}

\maketitle

\bigskip

A discrepancy of $3.3\sigma$ still remains between experiment and the standard model (SM) 
prediction for the muon anomalous magnetic moment $a_\mu\equiv(g_\mu-2)/2$ \cite{PDG2018}. 
The calculations of two-loop contributions to $a_\mu$ due to the electroweak interactions of the SM 
were completed quite some time ago~\cite{KKSS1992, PPdeR1995, CKM1995, CKM1996} and 
recently the numerically evaluated results of full two-loop electroweak corrections were 
presented~\cite{INY2019}. For more references, see Ref.\cite{PDG2018}.  
Two-loop electroweak corrections are expressed as the form 
\be
a_\mu^{\rm EW}({\rm 2~ loop})=\frac{5}{3}\frac{G_\mu m_\mu^2}{8{\sqrt 2}\pi^2}\sum_i C_i \frac{\alpha}{\pi}~.
\ee

\begin{figure}[htbp]
\begin{tabular}{cc}
\begin{minipage}{0.33\hsize}
  \begin{center}
   \includegraphics[width=50mm]{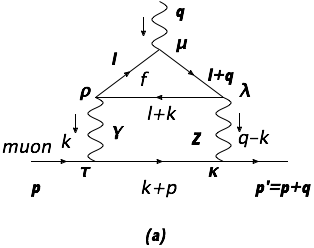}
  \end{center}
\end{minipage}%
 \begin{minipage}{0.33\hsize}
  \begin{center}
  \includegraphics[width=50mm]{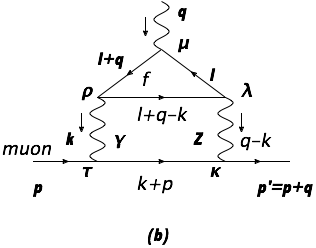}
  \end{center} 
\end{minipage}%
\end{tabular}
\caption{ Diagrams with fermion triangle loops connected to the muon line by a photon and a $Z$ boson. The diagrams with a photon and a $Z$-boson interchanged should be added.}
\label{FermionTriangle}
\end{figure}

Among the two-loop electroweak contributions there are an interesting subset of 
which are induced by triangle loops of charged fermions connected to the muon line by a photon and a $Z$ boson. The relevant  Feynman diagrams are shown  in Fig.\ref{FermionTriangle}. 
The fermionic triangle subdiagrams  in Fig.\ref{FermionTriangle} have the 
 Adler-Bell-Jackiw $VVA$ anomaly, which cancels out when all the fermions in each generation are included.  The analysis of this subset of diagrams was first made by Kukhto {\it et al.} \cite{KKSS1992} (KKSS). They used  a simplified version of the $Z\gamma \gamma$ vertex function by Adler \cite{Adler1969} and Rosenberg \cite{Rosenberg1963} for the fermionic 
 triangle subdiagrams and found that the loop contributions of leptons, i.e., $e, \mu$ and $\tau$,  were 
 enhanced by large logarithms of the form $\ln (m_Z/m_\mu)$ or $\ln (m_Z/m_\tau)$. Then followed 
 the studies of quark loop contributions~\cite{PPdeR1995, CKM1995}. 
%%%%%%%%%%%%%%%%%%%%%
%%%%%%%%%%%%%%%%%%%%%%%%%%%%%%%%%%%

In the calculation of two-loop electroweak contributions to $a_\mu^{\rm EW}$, the authors of \cite{CKM1995,CKM1996} employed the asymptotic expansion method and used the 't Hooft-Feynman gauge. Then in Ref.\cite{CM2017}, Czarnecki and Marciano, coauthors of  Refs.\cite{CKM1995,CKM1996}, illustrated the asymptotic expansion method in computing the triangle loop diagrams in Fig.\ref{FermionTriangle} with $f=$ top quark, since in this case there appear two large ratios of masses, $m_t^2/m_Z^2$ and $m_Z^2/m_\mu^2$. The result is made up of $\Delta C_Z$ and $\Delta C_G$ which are given in Eqs.(9) and (10), respectively,  of Ref.\cite{CM2017}. $\Delta C_G$ is the contribution of the neutral Goldstone boson.

Actually the top quark (more generally, a charged fermion) triangle loop contributions can be calculated exactly, 
 the results of which in the unitary gauge were already given  in Eqs.(13)-(15) of Ref.\cite{PPdeR1995}, in the form of  parametric representation with five Feynman parameters. In this letter I reexamine the contributions to $a_\mu$ from the  fermion triangle-loop diagrams depicted in Fig.\ref{FermionTriangle} in the unitary gauge and show that with use of the anomalous axial-vector Ward identity, the calculation in the unitary gauge {\it exactly coincides with} the one in the 't Hooft-Feynman gauge. The part which arises from the ordinary axial-vector Ward identity corresponds to the contribution of the neutral Goldstone boson. Then, for the top-quark contribution, the one-parameter integral form is obtained  up to  the order of $m_\mu^2/m_Z^2$.  The results are compared with those obtained by the asymptotic expansion method.

The muon state before and after the interaction with a photon field with momentum $q$ satisfies the following on-shell conditions:
\bea
{\overline u}(p+q)(\psl+\qsl)={\overline u}(p+q) m_\mu, \qquad \psl u(p)=m_\mu u(p), \label{OnShellCondition}
\eea
where $m_\mu$ is a muon mass. In the following calculation we put $q^2=0$ and  then the above on-shell conditions lead to  $p\cdot q=0$. 
It is well-known that the contributions of $VVV$ terms in the fermionic triangle subdiagrams  in Fig.\ref{FermionTriangle} (a) and (b) mutually cancel by  virtue of Furry's theorem 
while the $VVA$ terms have the Adler-Bell-Jackiw anomaly. Then we use the $Z\gamma \gamma$ vertex function derived by Adler \cite{Adler1969} and Rosenberg \cite{Rosenberg1963} for the fermionic triangle subdiagrams, which reads  in terms of the momenta shown  in Fig.\ref{FermionTriangle} as,
\bea
R^{\mu \rho \lambda}(q, -k, k-q)&=&\frac{1}{\pi^2}\int^1_0 dx \int^{1-x}_0 dy 
\Bigl[ m_f^2-x(1-x) k^2+2xy k\cdot q  \Bigr]^{-1}\nn\\
&&\times\biggl\{\Bigl[ x(1-x)k^2-xy(k\cdot q)  \Bigr]\epsilon^{\lambda\mu\rho q}
-xy(k\cdot q)\epsilon^{\lambda\mu\rho k}+xy~ q^\rho \epsilon^{\lambda\mu k q}\nn\\
&&\hspace{1cm}
-x(1-x) k^\rho \epsilon^{\lambda\mu k q}-y(1-y) q^\mu  \epsilon^{\lambda\rho k q}
+xy~ k^\mu  \epsilon^{\lambda\rho k q}  \biggr\} ~. \label{Full}
\eea
where $m_f$ is a fermion mass in the loop and $\epsilon^{\lambda\mu\rho q}=\epsilon^{\lambda\mu\rho \alpha}q_\alpha$, etc..   The convention 
$\epsilon^{0123}=-\epsilon_{0123}=1$ is used which agrees with Peskin and Schroeder 
\cite{PS1995} and not with Adler and Rosenberg \cite{Adler1969, Rosenberg1963} nor with KKSS \cite{KKSS1992}.  Note that the above expression is  the full-version of the $Z\gamma \gamma$ vertex function with $q^2=0$.
In Ref.\cite{KKSS1992},  KKSS used a simplified version  of $Z\gamma\gamma$ vertex function obtained from the full-version (\ref{Full}) by keeping,  in the numerator,  only the terms linear in 
the external photon momenta $q$ and discarding the $k\cdot q$ term in the denominator. The full-version 
$R^{\mu \rho \kappa}(q, -k, k-q)$ satisfies the electromagnetic current conservation, 
$q_\mu R^{\mu \rho \lambda}(q, -k, k-q)=0$ and $k_\rho R^{\mu \rho \lambda}(q, -k, k-q)=0$, and the anomalous axial-vector Ward identity
\bea
(k-q)_\lambda  R^{\mu \rho \lambda}(q, -k, k-q)
&=&-\frac{1}{2\pi^2}\epsilon^{\mu\rho k q}+
\frac{m_f^2}{\pi^2}\int^1_0 dx \int^{1-x}_0 dy 
\Bigl[ m_t^2-x(1-x) k^2+2xy k\cdot q  \Bigr]^{-1}\epsilon^{\mu\rho k q}~. \label{AxialVWI}
\eea
The first term is the Adler-Bell-Jackiw $VVA$ anomaly and independent of the  fermion mass $m_f$ in the loop, while the second term corresponds to the ordinary axial-vector Ward identity and is proportional to $m_f^2$. 

The calculation of the lower-loop in Fig.\ref{FermionTriangle} is performed by using
the $Z$-boson propagator in the unitary gauge, 
\bea
\frac{-i}{(k-q)^2-m_Z^2}\Bigl(g_{\lambda\kappa}-\frac{(k-q)_\lambda (k-q)_\kappa}{m_Z^2}\Bigr)~. \label{ZUnitaryGauge}
\eea
The contraction of $(k-q)_\lambda$ and the $Z\gamma \gamma$ vertex function gives 
the anomalous axial-vector Ward identity in (\ref{AxialVWI}). The anomaly term generates  a divergence in the loop integral. But it does not depend on the fermion and thus, in the SM, the contributions of the anomaly terms cancel out when all the fermions in each generation are included \cite{PPdeR1995}. Hence in the following we omit the anomaly term.  The calculation proceeds by making full use of an identity
\bea
i  \epsilon_{\mu\rho\nu\lambda}\gamma^\lambda \gamma_5=\gamma_\mu \gamma_\rho \gamma_\nu-g_{\mu\rho}\gamma_\nu-g_{\rho\nu}\gamma_\mu+g_{\mu\nu}\gamma_\rho~,
\eea 
and the on-shell conditions (\ref{OnShellCondition}). Discarding the terms with $\gamma_\mu$ and picking only those with $p_\mu$ and $q_\mu$, the remaining terms are found to be proportional to $(2p+q)_\mu$. Then the $y$-integration and the symmetrization in $x$ variable, i.e., $f(x)\rightarrow [f(x)+f(1-x)]/2$ are made. 

Finally the 
following expressions are obtained for the triangle loop contribution of a fermion $f$
 in the integral form with four Feynman parameters,
\be
C_{\gamma Z}(f)=N_c^f Q_f^2 I_{3f}\frac{12}{5}\Bigl\{A_{\gamma Z}(f)
-\lambda_fB_{\gamma Z}(f)\Bigr\},\label{CgammaZ}
\ee
where
\bea
A_{\gamma Z}(f)&=&\int^1_0 dx \int^1_0 dz_4 \int^{1-z_4}_0 dz_3 \int^{1-z_4-z_3}_0 dz_2 \biggl\{\frac{a(2+3z_4)}{a\kappa z_4^2+az_2+\lambda_f z_3}-\kappa \frac{a^2 z_4^3}{[ a\kappa z_4^2+az_2+\lambda_f z_3 ]^2}\biggr\}\label{Part1},\\
\nn\\
B_{\gamma Z}(f)&=&\int^1_0 dx \int^1_0 dz_4 \int^{1-z_4}_0 dz_3 \int^{1-z_4-z_3}_0 dz_2 \biggl\{\frac{1+3z_4}{a\kappa z_4^2+az_2+\lambda_f z_3}-\kappa \frac{a z_4^3}{[ a\kappa z_4^2+az_2+\lambda_f z_3 ]^2}\biggr\}, \label{Part2}
\eea
with $a\equiv x(1-x)$, $\lambda_f\equiv m_f^2/m_Z^2$ and $\kappa\equiv m_\mu^2/m_Z^2$; $N_c^f$,  $Q_f$ and $I_{3f}$ are the color factor,  the electric charge and the third component of weak isospin of the fermion $f$, respectively, with $N_c^f=3(1)$ for quarks (leptons). These are exact results,  and $A_{\gamma Z}(\lambda_f,\kappa)$ and $B_{\gamma Z}(\lambda_f,\kappa)$ are equivalent to
${\cal F}[m_f^2/m_\mu^2, M_Z^2/m_\mu^2]$ in Eq.(14) and ${\cal G}[m_f^2/m_\mu^2, M_Z^2/m_\mu^2]$ in Eq.(15) of Ref.\cite{PPdeR1995}, respectively. 
Having the expressions given in (\ref{Part1}) and (\ref{Part2}), it is easy to perform further integrations w.r.t. the Feynman parameters $z_2$, $z_3$, and $z_4 $. 
The term $A_{\gamma Z}(f)$ comes from the $g_{\lambda\kappa}$ part of the 
$Z$-boson propagator (\ref{ZUnitaryGauge}). On the other hand, 
 the  $B_{\gamma Z}(f)$ term arises from the ordinary axial-vector Ward identity,  and thus  is multiplied by the factor $\lambda_f$  in (\ref{CgammaZ}). 
Due to this factor,  $B_{\gamma Z}(f)$ is only relevant for the case $f=$ top quark~\cite{PPdeR1995}. It is interesting to note that the expression of $A_{\gamma Z}(f)$ turns out to be  the same as the one given by KKSS in Eq.(4.10) of Ref.\cite{KKSS1992},  which was derived by using the simplified version of $Z\gamma\gamma$ vertex function.

Now consider the calculation of the same diagrams in Fig.\ref{FermionTriangle}  in  the 't Hooft-Feynman gauge. The $Z$-boson propagator in the 't Hooft-Feynman gauge is given by
\bea
\frac{-i}{(k-q)^2-m_Z^2}g_{\lambda\kappa}~,
\eea
which is the same form as the $g_{\lambda\kappa}$ part of the $Z$-boson propagator
in the unitary gauge in (\ref{ZUnitaryGauge}). Hence the contribution generated from the 
$Z$-boson propagator in the 't Hooft-Feynman gauge is expressed as $A_{\gamma Z}(f)$ in (\ref{Part1}). In addition we need to consider the contribution generated from the neutral Goldstone boson $G^0$. 
The relevant diagrams are obtained from Fig.\ref{FermionTriangle} (a) and (b) with 
replacement of the $Z$-boson propagator by that of $G^0$. Also the axial-vector couplings of the $Z$-boson are replaced by the pseudo-scalar couplings of $G^0$ 
to the fermion in the loop and the muon. 
The loop-integral of the fermionic triangle subdiagrams  in Fig.\ref{FermionTriangle} (a) and (b) where the axial-vector vertex $\gamma^\lambda \gamma_5$ is replaced by the 
pseudo-scalar vertex $m_f\gamma_5$ gives 
\bea
\frac{m_f^2}{2\pi^2}\int^1_0 dx \int^{1-x}_0 dy 
\Bigl[ m_t^2-x(1-x) k^2+2xy k\cdot q  \Bigr]^{-1}\epsilon^{\mu\rho k q}~,
\eea
which is just one half of the second term of (\ref{AxialVWI}) (an ordinary axial-vector Ward identity). Now attaching the pseudo-scalar vertex $m_\mu\gamma_5$ to the 
muon line, the lower-loop integral is made and the following exact result is obtained for the contribution generated from the neutral Goldstone boson $G^0$:
\bea
C_{\gamma G^0}(f)=N_c^f Q_f^2 I_{3f}\frac{12}{5}(-\lambda_f)B_{\gamma G^0}(f)~,
\eea
where
\bea
B_{\gamma G^0}(f)&=&\int^1_0 dx \int^1_0 dz_4 \int^{1-z_4}_0 dz_3 \int^{1-z_4-z_3}_0 dz_2 \biggl\{\frac{2}{a\kappa z_4^2+az_2+\lambda_f z_3}-\kappa \frac{2a z_4^2}{[ a\kappa z_4^2+az_2+\lambda_f z_3 ]^2}\biggr\}~.
\eea
The expressions of $B_{\gamma Z}(f)$ and $B_{\gamma G^0}(f)$ look different at first glance but actually they are equivalent. Take the difference and we see that it
vanishes after the integration w.r.t. the variables $z_2$, $z_3$ and $z_4$.  Now it is clear that the calculation in the unitary gauge {\it exactly coincides with} the one in the 't Hooft-Feynman gauge. The part which arises from the ordinary axial-vector Ward identity in the 
unitary gauge corresponds to the contribution of the neutral Goldstone boson. 

For the case $m_f^2 \gg m_\mu^2$,  the integrations of $A_{\gamma Z}(f)$ and $B_{\gamma Z}(f)(=B_{\gamma G^0}(f))$ w.r.t. the variables $z_2$, $z_3$, and $z_4 $ are easily made up to ${\cal O}(\kappa)$. The results are 
\bea
A_{\gamma Z}(f)&=& {\widetilde A}_{\gamma Z}(f)+\frac{\kappa }{\lambda_f}\int^1_0 dx~a\biggl\{\frac{5}{3} \frac{a}{a-\lambda_f} \log\frac{a}{\lambda_f}+\frac{17}{18}+\frac{5}{3}\log\kappa \biggr\}+{\cal O}(\kappa^2)~,\label{Approxpart1}\\
B_{\gamma Z}(f)&=&{\widetilde B}_{\gamma Z}(f)+\frac{\kappa }{\lambda_f}\int^1_0 dx~\biggl\{\frac{4}{3} \frac{a}{a-\lambda_f} \log\frac{a}{\lambda_f}+\frac{8}{9}+\frac{4}{3}\log\kappa \biggr\}+{\cal O}(\kappa^2)~,\label{Approxpart2}
\eea
where 
\bea
{\widetilde A}_{\gamma Z}(f)= \frac{3}{2}\int^1_0 dx\frac{a}{a-\lambda_f}\log\frac{a}{\lambda_f}~,\qquad {\widetilde B}_{\gamma Z}(f)=\int^1_0 dx\frac{1}{a-\lambda_f}\log\frac{a}{\lambda_f}~. \label{tildeAtildeB}
\eea
The expression of $\lambda_f {\widetilde B}_{\gamma Z}(f)$ appeared already 
in the literature \cite{BZ1990,CCCK2001,CCK2001,CM2017} as the 
integral representation for the Barr-Zee diagrams. 
In the case of top quark, and thus $N_c^t Q_t^2 I_{3t}=\frac{2}{3}$ and $\lambda_t\simeq 3.6$, we expand $\frac{1}{a-\lambda_t}$ as
\be
\frac{1}{a-\lambda_t}=-\frac{1}{\lambda_t}\Bigl\{ 1+\frac{a}{\lambda_t} +\frac{a^2}{\lambda_t^2} +\cdots\Bigr\},
\ee
and we obtain after the $x$-integration
\bea
N_c^t Q_t^2 I_{3t}\frac{12}{5}A_{\gamma Z}(t)&=&
\frac{1}{\lambda_t}\Bigl\{\frac{2}{3}+\frac{2}{5}\log \lambda_t+\frac{1}{\lambda_t}
\Bigl[ \frac{47}{375}+\frac{2}{25}\log \lambda_t  \Bigr]+\cdots
\Bigr\}+\frac{\kappa }{\lambda_t}\Bigl\{\frac{34}{135}  +\frac{4}{9}\log\kappa   \Bigr\}
+{\cal O}\Bigl(\frac{\kappa}{\lambda_t^2},\kappa^2\Bigr), \label{CtP1Asymp}\\
&&\nn\\
-N_c^t Q_t^2 I_{3t}\frac{12}{5}\lambda_t B_{\gamma Z}(t)&=&\Bigl\{-\frac{16}{5}-\frac{8}{5}\log \lambda_t-\frac{1}{\lambda_t}\Bigl[ \frac{4}{9}+\frac{4}{15}\log \lambda_t  \Bigr]+\cdots\Bigr\} -\kappa\Bigl\{\frac{64}{45}  +\frac{32}{15}\log\kappa   \Bigr\}+{\cal O}\Bigl(\frac{\kappa}{\lambda_t},\kappa^2\Bigr) .
\label{CtP2Asymp}
\eea
We see that the above equations (\ref{CtP1Asymp}) and (\ref{CtP2Asymp}) reproduce the results of the asymptotic expansion method,  $\Delta C_Z$ and $\Delta C_G$, which are given in  Eqs.(9) and (10) of Ref.\cite{CM2017}. 
In addition, the subleading ${\cal O}(1/\lambda_t^2)$ terms and the nonleading ${\cal O}(\kappa)$ terms 
are included, respectively, in (\ref{CtP1Asymp}) and (\ref{CtP2Asymp}), which may serve as another check on 
the asymptotic expansion method. 

For the  contribution of top-quark $C_{\gamma Z}(t)$, the integral form of $\frac{8}{5}\Bigl( {\widetilde A}_{\gamma Z}(t)-\lambda_t  {\widetilde B}_{\gamma Z}(t)\Bigr)$  gives~ -5.134 with $\lambda_t= 3.6$, while the use of the leading ${\cal O}(1)$ and subleading  ${\cal O}(1/\lambda_t)$ terms in the expansions in  (\ref{CtP1Asymp}) and (\ref{CtP2Asymp}), which is the result of the asymptotic expansion method, $(\Delta C_D+\Delta C_G)$, given in Eq.(11) of Ref.\cite{CM2017},   leads to ~-5.140. Agreement is excellent and we see that the asymptotic expansion method works just fine for the case of top quark triangle loop diagrams.
The integral form of ${\widetilde A}_{\gamma Z}(f)$ in (\ref{tildeAtildeB}) is still 
applicable to estimate $C_{\gamma Z}(f)$ for the contributions from  the loops of $b$ quark, $\tau$ lepton and $c$ quark. 
Or a formula ${\widetilde A}_{\gamma Z}(f)\approx (3/2)(-2-\ln \lambda_f)$ can be  used since $\lambda_f\ll1$ for 
$f=b, \tau, c$. On the other hand, for the light quarks, i.e., $u$, $d$ and $s$ quarks, there is the issue of how to properly treat their triangle loop diagrams \cite{PPdeR1995, KPdeRP2002, CMV2003}. 
For the contributions to $a_\mu$ from the muon and electron triangle  loop diagrams, we arrive at the same formulae given by KKSS (Eqs.(4.11) and (4.12) of Ref.\cite{KKSS1992}) since $A_{\gamma Z}(f)$ in (\ref{Part1}) is the same expression as the one derived by them.

In summary, the two-loop electroweak contributions to $a_\mu$ from fermion triangle diagrams connected to the muon line by a photon and a $Z$-boson depicted in 
Fig.\ref{FermionTriangle} can be calculated without approximations. It is shown that 
the calculation in the unitary gauge exactly coincides with the one in the 't Hooft-Feynman gauge. The part generated from the ordinary axial-vector Ward identity in 
the unitary gauge corresponds to the contribution of the neutral Goldstone boson in 
the 't Hooft-Feynman gauge. For the top-quark contribution, the one-parameter integral form is obtained  up to  the order of $m_\mu^2/m_Z^2$.  The results are compared with those obtained by the asymptotic expansion method. We see that numerically agreement is excellent. And yet still remains the discrepancy of 3.3$\sigma$ between experiment and theory for $a_\mu$.

\begin{acknowledgments}
I thank T. Uematsu for reading the manuscript and for various supports. 
\end{acknowledgments}

%%%%%%%%%%%%%%%%%%%%%%%%%%%%%%%%

\end{document}